\documentclass[prb,preprint]{revtex4-1} 

\usepackage{amsmath}  
\usepackage{amsfonts} 
\usepackage{graphicx} 
\usepackage{natbib}
\usepackage[label font=bf,labelformat=simple]{subfig}
\newcommand\beq{\begin{equation}}
\newcommand\eeq{\end{equation}}
\newcommand\beqa{\begin{eqnarray}}
\newcommand\eeqa{\end{eqnarray}}
\newcommand\bs{\begin{split}}
\newcommand\es{\end{split}}

\newcommand\lb{\label}

\begin{document}
\title{The upward-driven disk, a steadily forced chaotic pendulum}
\author{Leo Maas}
\affiliation{Institute for Marine and Atmospheric research Utrecht, Utrecht University, Princetonplein 5, 3584 CC Utrecht, the Netherlands, email: L.R.M.Maas@uu.nl}
\date{\today}

\begin{abstract}
An "upward-driven disk" is a novel mechanical device built from LEGO parts. A circular disk is suspended from the point where it is sandwiched between two wheels, making it free to oscillate as a pendulum, but the location of that suspension point on the disk changes with time due to a steady upward driving force applied by rotation of one of the wheels.  The pendulum can dynamically flip between hanging downward and being inverted.  Depending on the upward drive and the initial conditions, the disk can exhibit steady rotation, periodic motion, or chaotic motion (and some of these for the same drive). This device serves as an easy-to-visualize analog of chaotic phenomena in other physical systems. Most notably, the upward driven disk mimics a simplified version of the celebrated Lorenz equations that are frequently used to describe fluid convection.
\end{abstract}

\maketitle 

\section{Introduction}

This paper introduces a novel driven pendulum. While most driving mechanisms act periodically, either on a pendulum's center of mass, its suspension point, or its moment of inertia, this new pendulum is driven steadily\citep{aref2007toying,guemez2009toys,baker2008pendulum,blackburn1989driven,tea1968pumping}. However, we will see that this steady driving force can result not only in steady motion, but also in periodic  and chaotic motion. This complex motion occurs because, while the driving force is applied steadily to the same location in space, its location with respect to the circular disk pendulum changes as the disk swings and translates.

Section \ref{sec2} will discuss the construction of the upward-driven disk and the measurement of its motion. Section \ref{sec3} presents the disk's behavior when forcing is either absent or consists in  a small, intermediate or large drive. Section \ref{sec4} will introduce the equations of motion, and approximations used.  Section \ref{sec5} will compare observed and simulated motion of the disk's center-of-mass for a few cases in which the strength of the upward drive is varied. It will address circumstances in which the disk, apart from exhibiting periodic and chaotic responses, also displays a steadily-rotating behavior. 
The  modifications of the equations governing the motion of the disk, required to capture this behavior, suggest 'missing physics' in the classical description, which we hypothesize is related to the role of friction. Section  \ref{sec6}  ends the paper by discussing the relevance of the upward-driven disk in comparison to other chaotic phenomena like fluid convection. 

\begin{figure}[htb]
\includegraphics[width=\textwidth]{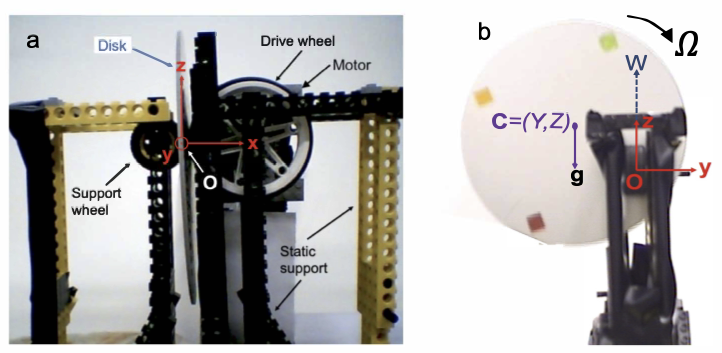}
\caption{(Color online) (a) Side view  and (b)  front view  of the upward-driven disk. The disk is sandwiched between two rotating wheels at O, the origin of a Cartesian $\mathsf{(x,y,z)}$  coordinate frame, fixed with respect to the lab. The big drive wheel on the right in (a) is set into clockwise rotation by the motor behind it, driving the disk upward at speed $W$. Gravity ${\bf g}$ provides a torque on the disk's center C, at time $T$ located at $\mathsf{(y,z)}=(Y(T), Z(T))$. The disk rotates at angular velocity $\Omega(T)$, which is positive when rotating clockwise. }\lb{UDD}
\end{figure}

\section{Experiment}\lb{sec2}

\subsection{Mechanical construction of upward-driven disk}\lb{subsecA}
The upward-driven disk, displayed in Figure~\ref{UDD}, is supported and driven by a  construction made of LEGO whose base, invisible in this figure, is put onto a flat table.  It consists of a vertically-hanging plastic disk of  $\mathcal{R} =8 $ cm radius that is sandwiched between two wheels, a drive wheel and a support wheel. The wheels are carried by two pillars that are put on the base. These wheels rotate about fixed horizontal axes. The point where the wheels contact the disk acts as the pivot point for the disk-pendulum; the disk  can rotate about that point. The wheel axes are 18 cm above the table surface so that the disk will never contact the surface below. The wheels have rims made of rubber that provide good traction, so the disk does not slip down. The LEGO construction has some flexibility, allowing the 1.5 mm thick disk to be inserted between them. While  slipping will never occur during disk translation, slipping will nevertheless play a role during disk rotation as the contact `point' of wheels and disk necessarily has some small yet finite extent.   The pressure exerted by the wheels on the disk needs to be adjusted empirically, so that the disk neither slides down nor loses its ability to rotate. This adjustment consists in having some weights leaning against the support pillars carrying the two wheels.  

The drive wheel's outer surface is a rubber O-ring of radius 3 cm and cross-sectional diameter of 3 mm. It rotates at a rate $f$ that can be varied in the range of 0.3 to 3 $rad/s$ by a Technic LEGO motor   (the grey structure behind the drive wheel in Figure~\ref{UDD}a). The motor is powered by a tuneable 10 Volt DC power supply (the LEGO 8878 Power Functions Rechargeable Battery and 8887 Transformer). 
The support wheel rotates freely, and friction prevents its point of contact with the disk from slipping. Ideally the rubber rims of both wheels should be toroidal to minimise the contact area, but in order to prevent out-of-plane wobble of the disk, one of the wheels has a 13 mm wide flat face, see Figure~\ref{UDD}b. 

The precise mass of the disk (here approximately 33 gram) is immaterial  to the outcome of the experiment, provided it is much smaller than the combined mass of the supporting structure, motor and power supply, so that the support does not sway when the disk moves. Due to the insensitivity to the mass, all quantities in the paper are "per unit mass", so that, for example, the unit of angular momentum is $m^2$.\\

\subsection{Disk coordinates}\lb{subsecB}

In three-dimensional space  a solid object (like the disk) has five degrees-of-freedom: three coordinates to locate its centre-of-mass, C, and two angles that specify the object's orientation relative to C.  We will use an  $\mathsf{(x,y,z)}$ Cartesian coordinate frame,  that is fixed in the lab frame, with its origin at O, the pivot point of the disk. The  horizontal direction that connects the support and drive wheel centers defines the $\mathsf{x}$-axis (Figure~\ref{UDD}a). The $\mathsf{y}$ and $\mathsf{z}$-axes are chosen along the disk plane, in orthogonal horizontal and vertical directions, the latter antiparallel to gravity, see Figure \ref{UDDsketch}.

In this coordinate frame the disk centre C's  time-varying position is labeled by $(X,Y,Z)$. Assuming that the disk does not wobble, C's position along the x-axis is fixed at X = 0, which also implies that one of the two angles referred to above vanishes. \footnote{This assumption is in practice not always justified. Experimentally, at times some slight wobble (of about 10 degrees) takes place, but this  will be ignored in the theoretical model, discussed in section \ref{sec4}.} The disk's angular orientation relative to centre C is given by tracking a single fixed point on the disk, the red mark M in Figure~\ref{UDDsketch}. The line CM makes an angle $\phi$,  measured positive in clockwise direction  relative to the prolongation of OC, the blue line  extending OC in Figure~\ref{UDDsketch}.

Parts of our analysis will be more convenient if we transform the Cartesian coordinates $Y,Z$ of the center of mass into polar coordinates $R, \theta$, given  by  \beq (Y,Z)=R (\sin \theta, \cos \theta).\lb{coord}\eeq
Here  $R(T)=|OC|$ denotes the radial distance of the disk center from pivot O as a function of time $T$, while  $\theta(T)$ denotes the clockwise angle of  line OC relative to the vertically upward $\mathsf{z}$-direction. Choosing $\theta=0$ along the positive $\mathsf{z}$-axis respects the left-right symmetry imposed by the direction of gravity. 

The sum angle, $\beta= \theta+\phi$, gives the disk's orientation relative to a vertically upward-directed line through C. Angles $\phi$ and $\beta$, indicated in  Figure~\ref{UDDsketch}, suffice to indicate the disk's orientation relative to OC or the vertical respectively.  The time derivative of $\beta$, indicated by a dot, represents the disk's angular velocity, \beq \Omega\equiv \dot{\beta}=\dot{\theta}+\dot{\phi}\lb{Xeq}.\eeq 

\begin{figure}[ht]\centering
\includegraphics[width=0.8\textwidth]{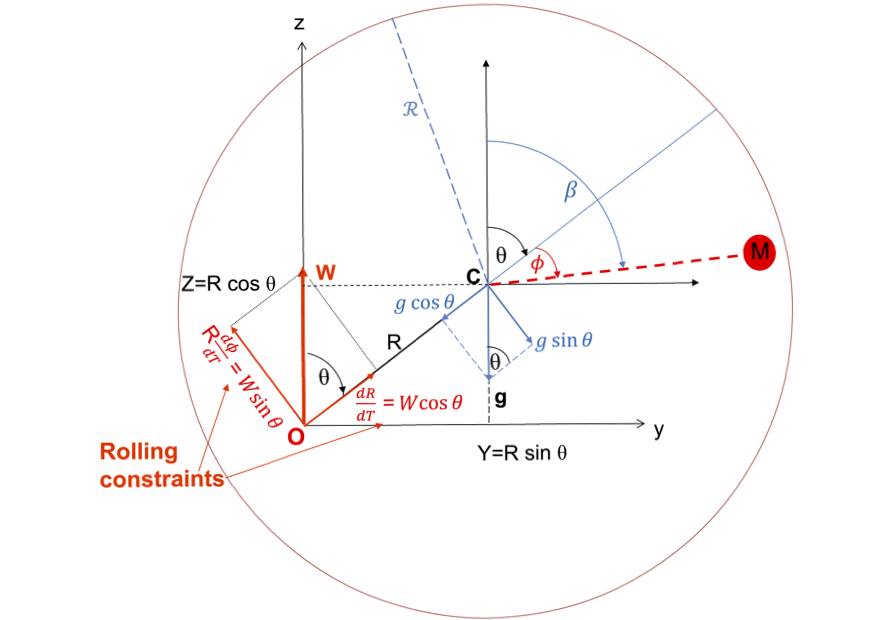}
 \caption{(Color online) Side view of an upward-driven, rotating disk. The disk of radius $\mathcal{R}$ is sandwiched in between two rotating wheels at O, the origin of a Cartesian $\mathsf{(y,z)}$ coordinate frame that is fixed in the laboratory. The wheels drive the disk upwards at a constant, tuneable velocity $W$. Gravity $({\bf g})$ provides a torque on the disk's center ${\bf C}=(Y(T), Z(T))$ which generates angular momentum $J(T)$. Note that  in this snapshot the disk appears as a pendulum in inverted position, having $Z>0$. The red mark M presents a fixed element of the disk whose orientation relative to the vertical is given by  angle $\beta =\theta+ \phi$.} \lb{UDDsketch}
\end{figure}

\begin{figure}[h]  
\includegraphics[width=\textwidth]{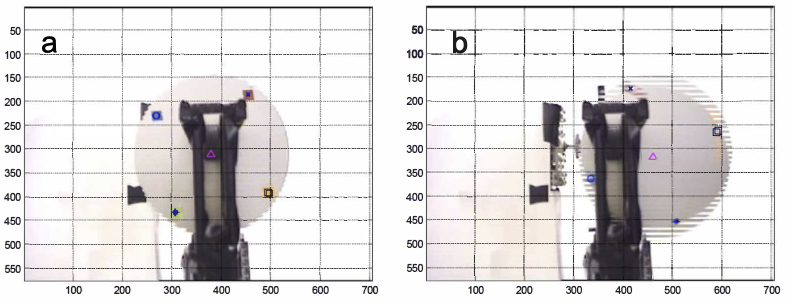}
\caption{{\small (Color online) (a) Image analysis determines the central pixel position of four square color marks near the rim of the disk, indicated by a symbol on top of each mark.  From these, the location of the center C, indicated by a purple triangle, is inferred. 
(b) Blurred image during rapid motion leading to poor determination of the centroid location of color marks.}\lb{center}}
\end{figure}

\subsection{Measuring Disk Coordinates}\lb{subsecC}

The moving disk is video-taped with a 550 $\times$ 700 pixel camera, and digitized video images are captured at a 25 Hz rate. Four square color marks (blue, green, red and orange) are glued diametrically onto the disk, see Figures~\ref{UDD}b and \ref{center}. With an RGB (Red-Green-Blue) image-analysis of each frame, the central pixel positions of the four colored areas are estimated.

Since errors obstruct  the accurate determination of each mark's location (for instance due to blurring when the disk moves rapidly, see Figure \ref{center}b), a minimization procedure is applied to these estimates that uses the fixed and known relative distances and orientations between the four marks. In Figure \ref{center} each mark's detected center is printed with different symbols on top of the color marks.  The pixel positions of the four color marks are subsequently converted to meters relative to the origin of the  coordinate system O,  calibrating the pixel to distance conversion ratio by using the measured disk size. 

Although the center of the disk and one of the four marks are frequently  hiding behind the support device (black construction in front of the disk in Figure \ref{center}) the remaining three will remain visible, which suffices for determining both the location of centre-of-mass C (indicated by a purple triangle in Figure \ref{center}) as well as the disk's orientation  $\phi$. 

Since the disk's mass $m$ is arbitrary, and for a solid disk appears both in its angular momentum $J$ as well as in  its moment of inertia $I$,  mass divides out in the angular momentum equation and in the following we will therefore express these quantities per unit mass as $\mathcal{J}\equiv J/m$ and $\mathcal{I}\equiv I/m$ respectively. The total angular momentum with respect to the origin O is,  \beq \mathcal{J}=\mathcal{I}\dot{\theta}+\mathcal{I}_C \dot{\phi}=(\mathcal{I}_C +\mathcal{I}_O)\dot{\theta}+\mathcal{I}_C \dot{\phi}.\lb{angl}\eeq 

By the parallel axis theorem, the moment of inertia per unit mass $\mathcal{I}$ for rotation around O is given by \beq \mathcal{I}= \mathcal{I}_C+ \mathcal{I}_O(T). \lb{moi}\eeq  This is the sum of  the moment of inertia of the disk rotating around center C, $ \mathcal{I}_C={1\over 2} \mathcal{R}^2$, which, recalling that $\mathcal{R}$ denotes the disk radius, is constant, and the moment of inertia $\mathcal{I}_O$ of the disk rotating around  pivot O at distance $R(T)$  that varies over time $T$. Per unit mass, the latter is given by \beq  \mathcal{I}_O= R^2\equiv Y^2+Z^2.\lb{IO}\eeq Hence, with (\ref{Xeq}),
angular momentum per unit mass can also be written as \beq \mathcal{J}= \mathcal{I}_O \dot{\theta}+I_C \Omega = \mathcal{I}\Omega - \mathcal{I}_O \dot{\phi}.\lb{ams}\eeq \\

\subsection{The disk's motion}\lb{subsecD}

When the drive is absent ($W=0$), the disk  operates as a standard damped pendulum. Its effective length $R=|OC|$ is then fixed,  see Figure~\ref{UDD}b. As with any pendulum, when $Y\ne 0$, the gravitational force produces a torque on the disk. Whenever it surpasses the static frictional torque, this  torque drives the disk  towards the downward-hanging, stable position, where $Z<0$.  

However, when the drive is on ($W > 0$) the point on the disk that is in contact with the wheels, i.e. coinciding with O, changes in time. Therefore, so do pendulum  length $R$ and angle $\theta$.    This happens also to a disk that initially has no horizontal misalignment  ($Y=0$) and is driven upwards from a stable position $(Z<0)$. As soon as  C rises above O, the disk behaves like an inverted pendulum that is at its unstable equilibrium point. Any noise causes the disk to  fall sideways and downwards. Like the standard (non-inverted) pendulum, the disk then engages in  oscillations about the downward-hanging, stable position. Although these oscillations will subsequently be damped by friction,  as the disk is   simultaneously  driven upwards ($W>0)$, its centre-of-mass again moves towards an unstable inverted position ($Z>0$)  until the disk falls over once more and a new cycle  commences. This goes on indefinitely. But does the disk's center  follow the same path during each cycle? We will look into this question, both experimentally and in a matching model description.\\

\section{Results}\lb{sec3}

\subsection{Zero Drive}\lb{subsec3A}
\begin{figure}[h]
\includegraphics[width=\textwidth]{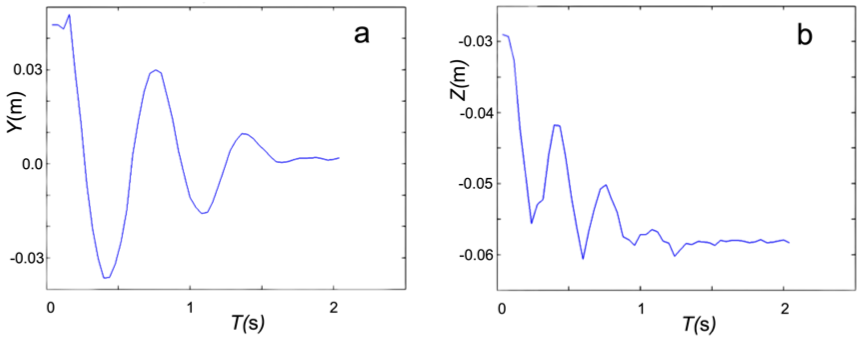}
\caption{The disk's center-of-mass (a) lateral  $Y$, and (b) vertical   $Z$  coordinate as a function of time $T$ for zero drive, $W=0~cm/s$.  }\lb{zero} 
\end{figure}

We initially carried out an experiment with no upward drive, $W=0$. Results are shown in Figure~\ref{zero}.  The disk behaves as a  physical pendulum that is strongly damped, but not overdamped, with the motion dying out after 3--4 oscillations.  Below in Section \ref{sec5zd}, we propose a damping model that fits this experimental data.

\subsection{Small Drive: Chaotic Regime}\lb{subsec3B}

The behavior of the disk for a small value of the upward drive, $W=1.4$ cm/s, starting from rest as an inverted pendulum, $Y(0)\approx 0$, $Z(0)= 4$ cm,  is shown in some detail in Figures~\ref{YZ14}-\ref{polar}  (and in Supplemental Material, Video 1).  
Initially, when the drive is turned on, the disk still creeps up a little further while the gravitational  torque pushes it over to fall to the left $Y<0$. It then drops down rapidly and shows a number of oscillations whose amplitude decays subsequently. When these oscillations have died out and its center C lies nearly straight below origin O, meaning $Y \approx 0$, see Figure \ref{YZ14}a, as time progresses, the disk   slowly creeps upwards. The constant slope displayed in each cycle by the disk's vertical position $Z$ as a function of time $T$, Figure \ref{YZ14}b, shows the disk moves upward at the nearly constant speed $W$ imposed by the drive.  \\

\begin{figure}[t] \includegraphics[angle=0,width=\textwidth]{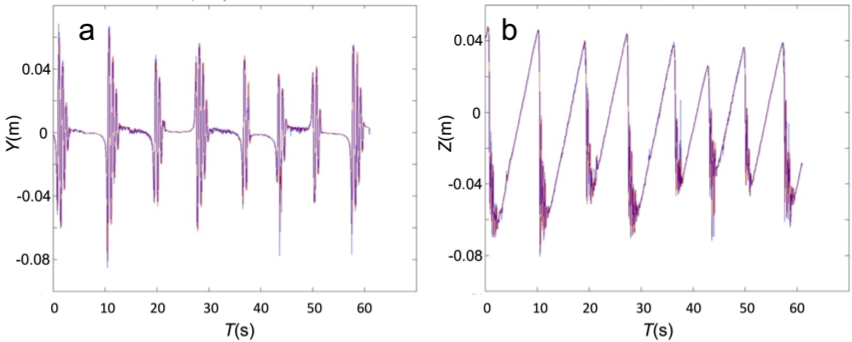}
\caption{(Color online) The disk's center-of-mass (a) lateral  $Y$, and (b) vertical   $Z$  coordinate as a function of time $T$ for upward-drive $W=1.4~ cm/s$. Blue and red lines represent raw and smoothed measurements respectively.}\lb{YZ14} 
\end{figure}
\begin{figure}[h]
\includegraphics[angle=0,width=\textwidth]{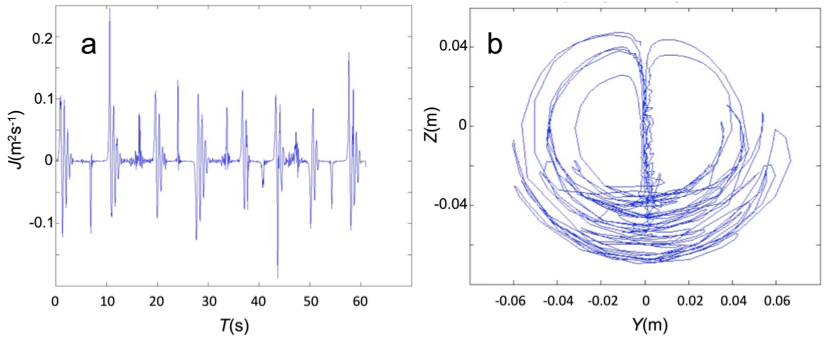}
\caption{The disk's (a) angular momentum $\mathcal{J}$ per unit mass as a function of time $T$ and (b) trajectory of its center C, located at $(Y,Z)$,  in the $\mathsf{y,z}$-plane  for upward-drive $W=1.4$ cm/s.}\lb{slow} 
\end{figure}

Even when the disk starts out precisely at the symmetry axis $\mathsf{y=0}$ and the gravitational torque vanishes, any misalignment with the symmetry axis produced either by noise or by a tiny residual displacement, causes a gravitational torque. When this torque surpasses the torque caused by static friction it generates angular momentum $\mathcal{J}$ (Figure \ref{slow}a) while pushing the disk sideways, Figure \ref{YZ14}a. Because gravitational torque is proportional to $Y$, the smaller $Y$ is initially, the higher the disk can be displaced upwards before this torque overcomes static friction and the deeper it falls down when eventually overturning, see the differing heights reached by the disk in successive cycles (Figure \ref{slow}b). During this phase, friction lends the disk some metastability. In its following stable phase ($Z<0$) oscillations decay until angular momentum is depleted. At some point, when nearly all angular momentum is lost, static friction arrests the disk's rotary motion. While still nonzero, $Y$ can then become very small, defining the disk's `new' initial $Y$ in the next cycle (Figures \ref{YZ14}a and \ref{slow}b). 

\begin{figure}[t]
\includegraphics[angle=0,width=\textwidth]{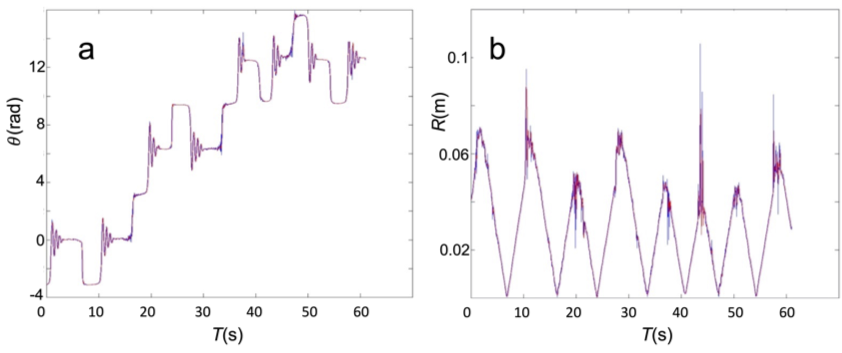}
 \caption{(Color online) The disk's (a) angle $\Theta =\theta-\pi$ with respect to gravity ($\Theta=0\; \mbox{mod}\; 2\pi$) and (b)  pendulum length, $R = |OC|$,  as a function of time $T$ for upward-drive $W=1.4~ cm/s$.}\lb{polar}
\end{figure} 
The motion of the center-of-mass C can also be nicely seen in polar coordinates, Figure \ref{polar}. Smooth, rapid jumps in $\theta$ by $\pm \pi$ occur when C passes O during its ascent, see e.g. near $T=7.5$ s in Figure \ref{polar}a. This smooth jump to an inverted position is eventually followed by another rapid jump by $\pm\pi$ (such as near $T \approx 11$ s), representing the descent phase, immediately followed by pendulum oscillation.  During subsequent cycles, once  the disk is in inverted position, it may fall repeatedly to the same side, C obtaining the same sign($Y$) as that during its previous passage of O (see $T \approx 15$ s). This leads to successively larger or smaller values of $\theta$, reaching plateaus at multiples of $2 \pi$, see Figure \ref{polar}a. But, when the disk alternates sides, $\theta$ returns to the same plateau, as for instance seen in Figure \ref{polar}a near $T \approx 11$ s, where $\theta$ returns to near zero.
Figure \ref{polar}b also shows that for low drive, the disk's ascending "slow" phase (visible as a rectilinear decrease or increase of $R(T)$) takes much longer than its descending "fast" phase, when it executes oscillatory motion (visible near $R$'s maxima). These phases can also be clearly distinguished in $Z(T)$, Figure \ref{YZ14}b. Note that $R$'s maxima represent both its maximum upward displacement, its radius during its descent  following a near-circular  path, as well as the start of its oscillations around its  stable downward position.
The unusually large spikes near these maxima in Figure \ref{polar}b are caused by limitations in the camera frame rate to capture the rapid motion during the fast descent.

\begin{figure}[t]
\includegraphics[angle=0,width=\textwidth]{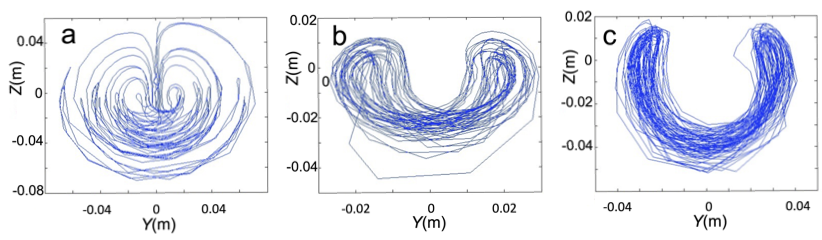}
  \caption{The disk's center-of-mass coordinates in the vertical plane $Y,Z$ (m) for upward-drive $W=$ (a) 3.9, (b) 5.5 and (c) 7.1~ cm/s.}\lb{YZs} 
\end{figure}
\subsection{Intermediate and Large Drive: Chaotic, Periodic and Steadily Rotating Regimes}\lb{subsec3C}

When the upward-drive speed is cranked up, one loses the clear separation between the slow and fast phases seen in the previous section. The  disk's motion remains chaotic  at an upward drive speed $W=3.9$ cm/s, see Figure~\ref{YZs} (and Supplemental Material, Video 2). But at  higher speed, $W=5.5$ cm/s, chaos seems to be suppressed, Figure~\ref{YZs}b, while at $W=7.1\;cm/s$ it appears completely eliminated, see Figure~\ref{YZs}c (and also Supplemental Material, video 3, at still higher speed $W=8.8$ cm/s). 

Surprisingly, the disk also shows the presence of two symmetric, steadily rotating states (see Supplemental Material, Video 4).  In these modes, the center-of-mass  is stationary. It is located a bit off-center $(|Y|>0)$, at mid-level $Z=0$, while the disk rotates at a fixed speed $| \Omega| =$ constant $>0$, either in clockwise or anticlockwise direction. These modes cannot be reached from rest. Rather one has to start at an initially large  upward-drive speed $W$, give the disk some initial rotation, and keep the disk's center C a little sideways from the pivot O where it is sandwiched. The disk will then be trapped in this steadily rotating state. When left undisturbed, the disk can stay spinning for over an hour. However, perturbing the disk in such a steadily rotating regime, for instance by touching it, brings it easily  out of equilibrium, so that it returns either to a periodic or chaotic state. This testifies that multiple equilibria may exist simultaneously, i.e. for the same drive. When one refrains from perturbing this steady rotation of the disk mechanically, but instead gradually reduces its upward-drive,  these equilibria eventually turn unstable too, as if there exists a lower bound on $W$ below which these equilibria destabilize. 

The observant reader may notice that the apparatus used in Video 4 of the Supplementary Material is different from the one in Videos 1-3. The first apparatus was damaged during transport, and we are not certain whether it would have been able to exhibit the steadily rotating motion that is shown in this video using that original apparatus. Achieving this state requires careful adjustment of the normal force of the wheels on the disk, which was possible in the second apparatus by turning a screw that adjusted the force.

\section{Theoretical Model of Upward-Driven Disk}\lb{sec4}

The equations governing the disk's motion consist of two kinematic equations describing the upward motion of the disk at its point of contact and one dynamical equation describing the evolution of its angular momentum. 

\subsection{Kinematic Equations}\lb{subsecKE}
The kinematic equations follow from the no-slip (rolling) constraint that applies at the contact points of disk and wheels. 
Projecting the constant upward drive $W$ at the point of contact onto radial and azimuthal directions reveals that the drive changes both the  disk center's radial position $R$ as well as its azimuthal orientation $\phi$ (see Figure~\ref{UDDsketch}):
\begin{subequations}\lb{eq:angular}
\begin{align}
\dot{R}&=W \cos\theta,\lb{eq:angular1}\\
R\dot{\phi }&=W \sin \theta,\lb{eq:angular2}\end{align}
\end{subequations}

To check their validity, let's look at two limiting cases: (1) $W=0$ and (2) $g=0$. In the first case, $W=0$, equations (\ref{eq:angular}) state that the distance $|OC| \equiv R$ is constant and that disk orientation $\phi$, relative to line OC, is constant. This is clearly what one should expect for a disk having a fixed point at O:  while line OC  changes its orientation $\theta$ due to the gravitational torque, mark $M$ just rotates with it, keeping $\phi$ fixed.  In the second case, $g=0$, a drive $W > 0$ moves the disk upwards, but in the absence of a torque the disk does not rotate with respect to the lab. In this case, the $Y$-coordinate of the disk's center C remains constant, as does its orientation relative to the vertical $\beta$, while $\dot{Z} = W$. Using $(Y_0,Z_0)$ as the initial position of the center, write $R^2 = Y_0^2 + (Z_0+ WT)^2$, then take its time derivative and substitute $ Z_0+WT=R\cos \theta$ to yield 
 Eq. (\ref{eq:angular1}).
Then take the time derivative of $R \sin \theta = Y_0$ and substitute Eq. (\ref{eq:angular1}) which yields Eq. (\ref{eq:angular2})
using the fact that $\beta =$ constant implies $\dot{\phi}= -\dot{\theta}$.

With the definition of total angular velocity $ \Omega$ in (\ref{Xeq}) as the sum of the rate $\dot{\theta}$ with which the disk center C rotates around pivot O and the rate $\dot{\phi}$ with which the disk rotates around its own center C, the kinematic constraint equations (\ref{eq:angular}) lead to the rate of change of the center-of-mass position vector $(Y,Z)$. Employing (\ref{coord}) and (\ref{Xeq}), while multiplying (\ref{eq:angular1}) by $\sin \theta$ and subtracting $\cos \theta$ times (\ref{eq:angular2}), and similarly, multiplying (\ref{eq:angular1}) by $\cos \theta$, adding $\sin \theta$ times (\ref{eq:angular2}), gives respectively
\begin{subequations}\lb{eq:YZeqs}
\begin{align}
\dot{Y}&=Z \Omega,\lb{eq:YZeqs1}\\
\dot{Z }&=-Y  \Omega+ W.\lb{eq:YZeqs2}\end{align}
\end{subequations}

The forcing here appears as a nonholonomic constraint in (\ref{eq:YZeqs2}), i.e. a constraint on the velocity acting on the contact point of wheels and disk.
Remarkably, in this dynamic set-up, in which the disk is simultaneously pulled down by gravity and pushed up by the drive wheel, the angular velocity pushing the centre-of-mass turns out to be given by the total angular velocity $\Omega$. 

\subsection{Angular Momentum Equation}\lb{subsecAM}

The downward-directed gravitational force per unit mass ${\bf g}$ acting at disk center C can be decomposed into a torque acting in azimuthal ($\theta)$ direction, and a radial force  that points from C to O -- see  its components displayed as blue vectors in Fig~\ref{UDDsketch}.  At O this radial force is balanced by static friction between wheels and disk. When some slipping occurs, for $|\theta|<\pi/2 $, this radial force pushes the disk down and towards the symmetry line $\mathsf{y}=0$; for $|\theta|>\pi/2 $ it pulls the disk down and away from this symmetry line, while this radial force is absent when $\theta= \pm \pi/2.$ The downward directed component of this force in O  is however smaller than the motive\cite{besson2007teach}, upward directed frictional force, producing the steady upward rolling motion of the disk  at the point of contact. In Section \ref{sec5} we hypothesize that this component may nevertheless play a role explaining the disk's unexpected steadily rotating mode. 

The rate at which total angular momentum with respect to O changes is determined by the torques acting on the disk: the gravitational torque, $gY$, and a retarding frictional torque, $-k \Omega$. The latter is  modelled as linearly proportional to the total angular velocity $\Omega=\dot{\theta}+\dot{\phi}$ as the disk both rotates around O and spins around C. The proportionality of  torque to velocity can be attributed mainly to drag by the air through which the disk moves\cite{squire1986pendulum,taylor2005classical,baker2008pendulum}. As seen in the next subsection, its presence explains why, without any forcing, a disk eventually reaches a location below C,  at $Y \approx 0, Z<0$, regardless its initial state.  

Starting with (\ref{ams}),  and using (\ref{coord}) and (\ref{eq:angular2}),  the total angular momentum per unit mass can be rewritten  as 
\beq \mathcal{J}= \mathcal{I}\Omega - R^2 \dot{\phi}=\mathcal{I} \Omega-RW\sin \theta=\mathcal{I} \Omega-WY,\lb{angmom2}\eeq where the last identity follows from (\ref{eq:angular2}),  (\ref{Xeq}) and (\ref{coord}). 

Thus the angular momentum  equation is
\beq {d \mathcal{J}\over dT }={d\over dT} (\mathcal{I}  \Omega-WY)= gY - k \Omega.\lb{Jevol}\eeq
Employing (\ref{eq:YZeqs1}), and using that $dR^2/dT=2WZ$, which follows from (\ref{eq:angular1}) and (\ref{coord}), this reduces to  
\beq \mathcal{I} \dot{\Omega}=gY-k  \Omega-WZ \Omega,\lb{angmom4}\eeq
where frictional torque is modelled as $-k\Omega$.
Together with (\ref{eq:YZeqs}), this equation will be used when comparing observed to modelled disk motions in Section \ref{sec5}.
Note that  if one were to neglect $-k \dot \phi$ in the expression for  frictional torque $-k \Omega$, Eq. (\ref{angmom4})  would yield an erroneous result in case gravity vanishes. When  $g=0$ the disk does not rotate around its centre, hence $\beta=const$, so that  $\Omega=0$, which implies that all terms in (\ref{angmom4}) should vanish. However,  when   frictional torque is modelled as $-k \dot{\theta}$, this  is nonzero since angle $\theta$ changes for C steadily moving upward in positive $\mathsf{z}$ direction. 

If $g \ne 0$ and $Y$ is very small, as when the disk is steadily driven "straight" up for a case where $W$ and $Z$ are small,  $\Omega $ might be small too. But  both $Y$ and $\Omega$ must nevertheless be nonzero as the drift of $Y$, observed in  Figure \ref{YZ14}a, suggests.
\subsection{Simplifications}\lb{subsecSIM}
The angular momentum equation (\ref{Jevol}) simplifies when we assume that both the moment of inertia per unit mass $\mathcal{I}$ as well as the friction coefficient $k$ are constant.  This is justified when both the maximum displacement of the disk's center is much less than its radius $R(T) \ll \mathcal{R}$, and when the angular velocity is always very large, such that $k_2 |\Omega | \gg k_1$ and, according to (\ref{kt}), $k \rightarrow k_0$, a constant.  This  corresponds to the large drive regime. The former condition implies $\mathcal{I}_O \ll \mathcal{I}_C$, such that with $\mathcal{I}\approx \mathcal{I}_C$, the angular momentum in (\ref{angl}) simplifies to  $\mathcal{J}\approx \mathcal{I}_C \Omega$.  The angular momentum equation (\ref{Jevol})  then reduces to 
\beq \mathcal{I}_C \dot{\Omega}=  gY -  k_0 \Omega.\lb{am}\eeq While these assumptions are not necessarily respected in the physical experiments, the mathematical simplifications in the model  make its consideration worthwhile. Equation (\ref{am}), together with (\ref{eq:YZeqs}), effectively lead to a one-parameter set of three coupled equations. By appropriate choice of length ($L=2k_0^2/(g\mathcal{R}^2)$) and time $(\tau=\mathcal{R}^2/(2k_0)$) scales,  constants $g,\; \mathcal{I}_C$ and $k_0$ can all be scaled out\citep{strogatz2001nonlinear}, where $\mathcal{R}$ represents the disk radius. This leaves a set of three dimensionless equations in which, apart from  initial location and angular velocity, only a single parameter appears, a dimensionless  drive $\mathsf{w}=\tau W/L$.  Expressing dimensionless variables in corresponding lower-case symbols
\beq  T=\tau t,  \Omega= \omega/\tau, Y=Ly, Z=Lz  \lb{scales} \eeq
 the nondimensional equations yield  
\begin{subequations}\lb{eq:DLEQ}
\begin{align}
\dot{\omega}&=y-\omega,\lb{eq:DLEQ1}\\
\dot{y}&=\omega z,\lb{eq:DLEQ2}\\
\dot{z}&=-\omega y+\mathsf{w}.\lb{eq:DLEQ3}
\end{align}
\end{subequations}
Here, a dot now refers to a  dimensionless time ($t$) derivative, $\omega$ represents the angular velocity, while $y$ and $z$ denote the horizontal and vertical coordinates of disk centre C in a dimensionless Cartesian $\mathsf{(y,z)}$-frame of reference.  

When forcing is absent $(\mathsf{w}=0)$, equations (\ref{eq:DLEQ}) describe a  damped physical pendulum. Equations (\ref{eq:DLEQ2}, \ref{eq:DLEQ3})  imply that in this case pendulum length $r =\sqrt{y^2+z^2}$, the distance between pivot O and disk center C,  is constant. Hence, the center-of-mass C follows a circular path and its coordinates can be described by polar angle $\theta$, measured positive in clockwise direction from the positive $z$-axis, \beq (y,z)=r(\sin \theta,\cos \theta). \lb{coo}\eeq
Inserting these into (\ref{eq:DLEQ2}) or (\ref{eq:DLEQ3}) shows that in this unforced case, angular velocity
\beq  \omega = \dot \theta. \lb{av}\eeq
Inserting (\ref{coo}) and (\ref{av}) into (\ref{eq:DLEQ1}) and replacing $\theta=\pi+\Theta$, to ensure $\Theta=0$ corresponds to the rest position in which $z<0$,  yields 
\beq \ddot{\Theta}+\dot{\Theta}+r \sin \Theta =0, \lb{claspend}\eeq the dimensionless equation describing an unforced and damped physical pendulum \citep{baker2008pendulum,taylor2005classical,strogatz2001nonlinear}. 

Conventionally,  forcing of this pendulum appears   as a nonzero right-hand side in (\ref{claspend}), in particular in the form of a periodic (sinusoidal) forcing\cite{baker2008pendulum,blackburn1989driven}. This torque, periodically alternating in direction, is responsible for aperiodicity in its response. If this forcing is replaced by a time-independent torque, depending on its strength, due to the damping, the pendulum either oscillates periodically around a steadily rotating state or gets trapped at a fixed angle\citep{strogatz2001nonlinear}, but it will not show any complex response. It reduces our third order system (\ref{eq:DLEQ}) to a planar second order system, which, by the Poincar\'e-Bendixson theorem, no longer supports chaos\cite{coddington1956theory}.  

In our dimensionless system, such type of forcing  would enter as a time-periodic or steady term in the right-hand of (\ref{eq:DLEQ1}). However, instead employing a forcing in (\ref{eq:DLEQ3}) for nonzero drive, as we have in the upward-driven disk, is  unconventional. Moreover, a steady drive still leading to an aperiodic response, as we find  experimentally, is  unexpected and will be investigated here.

When $\mathsf{w}>0$,  pendulum length $r$ is no longer constant, and equations (\ref{eq:DLEQ}) are identical to a simplified, one-parameter version of the  celebrated Lorenz equations \citep{Lorenz1963,maas1994simple}. The three Lorenz equations  rudimentarily describe convection in a fluid heated from below. Solutions of the Lorenz equations are known to possess a spectacular sensitive dependence  on their initial conditions, displaying a strange attractor and what has been commonly referred to as `chaos'. The  one-parameter version   is obtained in the limit that viscous damping dominates over diffusion. For this reason the steadily-forced pendulum equations (\ref{eq:DLEQ}) are referred to as  the diffusionless Lorenz equations (DLE)\cite{VanderSchrier2000,maas2004theory,wei2014new,valani2024infinite}. While the DLE have fewer terms than the Lorenz equations, they still  display the same rich phenomena, including periodic and chaotic states. When forced from different initial conditions, over some range of $\mathsf{w}$, these states can actually be simultaneously present for one and the same drive $\mathsf{w}$\cite{VanderSchrier2000,maas2004theory}. This predicted behavior is what was found in our actual experiments. 

The DLE  originally inspired the construction of the upward-driven disk. As we will find in the next Section  \ref{sec5}, in view of the similarity between its simulated and  observed behavior, the upward-driven disk can be argued to represent a purely mechanical model of convection.  Note that the dynamics of the center-of-mass which it describes also played a pivotal role in the Malkus waterwheel which it resembles\cite{malkus,matson2007malkus}. But the upward-driven disk lacks any real fluid motion, such as in the  Malkus waterwheel which consists of a set of leaky cups attached to the rim of a wheel that are continuously filled with water at the top-cup, its dynamics mimicking that predicted by the Lorenz equations \citep{malkus,strogatz2018nonlinear,matson2007malkus}.

\section{Analysis and Comparison of Measurements and Simulation }\lb{sec5}

A notable feature of this experimental system is that at times, the disk is swinging freely, while at other times the disk is not just motionless (with $Y$ precisely equal to zero), but stuck at some small but non-zero $Y$.  At these positions, the gravitational force $gY$ is perfectly balanced by friction.  By definition, since the disk is stopped, this must be static friction, sometimes termed `stiction.'  Stiction is notoriously difficult to accurately model numerically.  \citet{pennestri2016review}  has reviewed a number of comparable models, comparing their accuracy to the difficulty of implementation.  Generally, stiction is modeled as a speed dependent damping rate at low speeds.  The accuracy depends not just on how physically realistic the model is, but also on the numerical method and the behavior being modeled.  Here, we have used the zero drive data to determine an adequate model that also provides reasonable results for non-zero drive.  While we include some speculative comments about the physical nature of the friction based on the form of this model, this is not the main topic of this work.

\subsection{Zero Drive Behaviour} \lb{sec5zd}
\begin{figure}[h]
\includegraphics[width=\textwidth]{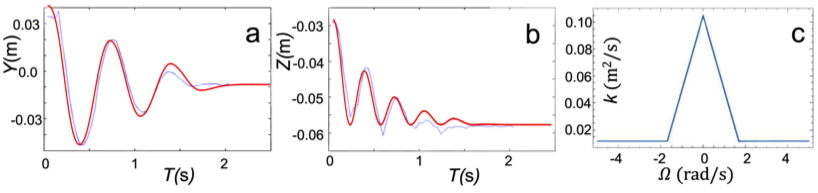}
 \caption{(Color online)   In the absence of a drive force ($W=0$), oscillations of the disk are observed (blue line) and modeled (red line) using  a drag force friction law $ - k \Omega$, where $k$ depends on angular velocity. Disk center C is initially located at $R(0)\approx 0.058 \;m$, $\;\theta(0) \approx 127^o$, having   initially vanishing angular velocity, $\Omega(0)=0$. (a) shows $Y(T)$, (b) $Z(T)$. (c) shows how the friction coefficient $k$  depends on angular velocity $ \Omega$.}\lb{kcompare} 
\end{figure}

Figure~\ref{kcompare} shows the zero-drive experimental data again in comparison to a numerically determined orbit used to study the damping behavior. Without forcing, disk orientation $\phi=$ constant and angular velocity simplifies to $\Omega=\dot{\theta}$.  The numerically determined orbit of the disk's center C was found as a  solution of the kinematic equations (\ref{eq:YZeqs}) and angular momentum equation (\ref{angmom4})  using a constant moment of inertia $\mathcal{I}_O$.

The frictional torque is modelled as $-k \Omega$.   We hypothesize that the linear dependence on angular velocity in this friction law represents the aggregate of viscous drag by the air through which the disk moves\citep{squire1986pendulum} as well as friction by the disk when slipping during rotation along the rubber wheels\citep{quiroga2017dynamics,cross2016coulomb}.    The decay of motion produced by an initial lateral displacement appears to depend nonlinearly on the disk's angular velocity $ \Omega$.  We find that the following expression for the friction coefficient,  \beq k= \max(k_1-k_2 | \Omega|, k_0), \lb{kt}\eeq 
 best fits the experimental data in Figure~\ref{kcompare}c. 
This form means that during the final stage of the damped oscillations (when $\Omega \rightarrow 0$ and $k_1>k_0$), motions are damped  faster than during the rapidly swinging phase. This is captured by the tuned friction coefficient  (\ref{kt}), shown in Figure~\ref{kcompare}c. Here, coefficients $k_0=0.012 ~m^2/s, ~k_1=0.103~m^2/s, ~k_2=0.055~ m^2$ have been determined  by fitting the observed evolution of the disk's centroid to a solution of the equation describing a damped physical pendulum --- see equation (\ref{claspend}) below.
 We speculate that in the beginning, for large angular velocities, the disk may  slip partially, which reduces friction, while at the final stage, at low angular velocities, friction increases. 

\subsection{Small Drive Behaviour}

\begin{figure}[ht]
\includegraphics[angle=0,width=\textwidth]{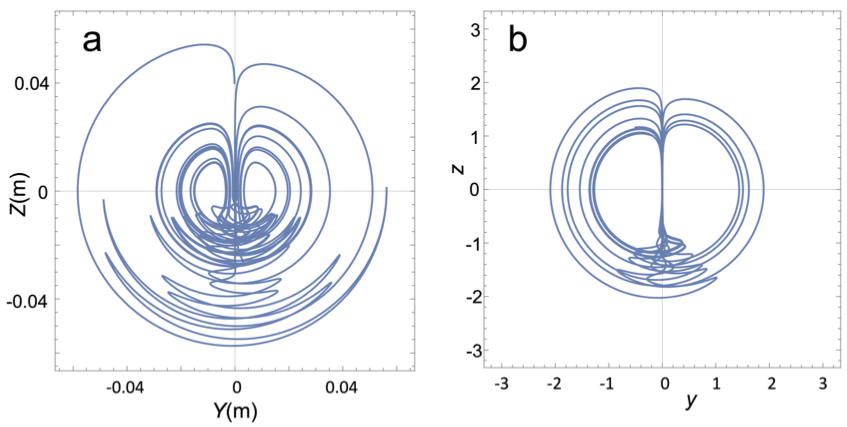}
  \caption{(Color online) (a) Orbit of the disk's center C in the vertical  $Y,Z$  plane computed using equations (\ref{eq:YZeqs}), (\ref{angmom4}) and  (\ref{kt})  for upward-drive   $W=$ 1.4 cm/s. (b) Orbit of the disk's center C in the dimensionless vertical $y,z$ plane  computed using equations (\ref{eq:DLEQ}),  for small upward drive $\mathsf{w}=0.1$, launched from initial location (-0.175,1). }\lb{UDDths} 
\end{figure}

Figure \ref{UDDths}a shows a numerically determined orbit of the disk's center C as  solution of the kinematic equations (\ref{eq:YZeqs}) and angular momentum equation (\ref{angmom4}) that incorporates the dynamic moment of inertia $\mathcal{I}_O$, see (\ref{moi}). Here, we use the state-dependent friction coefficient $k$, given in (\ref{kt}),  that was calibrated in Section \ref{subsec3A}.  We employ the same small upward drive $W=1.4$ cm/s used experimentally in Section \ref{subsec3B}. Given that we are looking at realisations of a chaotic pendulum, this figure resembles the observed motion of the disk's center, shown in Figure \ref{slow}b, fairly well. It displays similar oscillations in its downward hanging phase, a comparable slow upward ascent near the symmetry axis to comparable heights, and similar variability with regards to the sides to which its drops over once in inverted position. 

Despite the fact that the assumptions allowing the use of the DLE (\ref{eq:DLEQ}) do not really apply  for weak drive, numerical integration of that set for a small dimensionless drive $\mathsf{w}=0.1$, displayed in Figure  \ref{UDDths}b, shows qualitative agreement with that displayed in Figure \ref{slow}b as well. It correctly shows the depletion of angular momentum after reaching the stable position and its return to the symmetry axis $\mathsf{y=0}$  during each slow ascent phase. At this low upward drive, during the ascent phase, the vertical coordinate increases steadily over much of each cycle, see Figure~\ref{YZ14}b. This echoes the  presence of an (unstable) equilibrium state in the dimensionless DLE (\ref{eq:DLEQ}) given by $z=\mathsf{w} t+z_0, ~  \omega=y=0$, with $z_0=z(0)$,  in  experiments occasionally leading to the wheels ejecting the disk (see Supplemental Material, Video 2).  This state forms the relic of the motionless diffusive state in the original Lorenz equations\citep{Lorenz1963}. While unstable, it plays an important role in approximating the DLE (\ref{eq:DLEQ}) by an  iterated nonlinear  multi-cusped map\citep{VanderSchrier2000,maas2004theory}. The DLE can then  be integrated approximately during successive cycles, appropriate matching conditions ensuring continuity from one cycle to the next.  Due to the  loss of memory of its original lateral starting position,  $y(t_n)\ll1$, it is chaotic, where time $t_n$ denotes the beginning of the $n$-th cycle. 

\subsection{Intermediate and Large Drive Comparison}
For upward drives used in the experiments shown in Figure \ref{YZs}, a comparison is made between the disk's center-of-mass position C and that simulated  both by the full equations, (\ref{eq:YZeqs}) and(\ref{angmom4}),  as well as by the simplified set, represented by the DLE (\ref{eq:DLEQ}). This  comparison will be done both in physical $YZ$-space as well as in the three-dimensional $\mathcal{J} YZ$-phase space, which, interestingly, embeds physical space.  

\begin{figure}[h]
\includegraphics[angle=0,width=\textwidth]{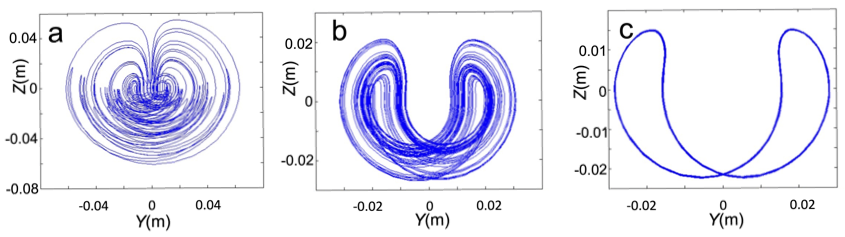}
  \caption{Orbits of the disk's center C in the vertical plane $Y,Z$  computed using equations (\ref{eq:YZeqs}), (\ref{angmom4}) and  (\ref{kt})  for upward-drive  (a) $W=$ 3.9, (b) 5.5, (c) 7.1 cm/s. }\lb{YZth} 
\end{figure}

When the drive increases, Figure~\ref{YZs} showed a transition from chaotic to periodic orbits. This  is substantiated by comparing these experimentally observed orbits to those computed with the theoretical model,  (\ref{eq:YZeqs}) and   (\ref{angmom4}).
For these settings the orbits simulated are shown in  Figure~\ref{YZth}. They appear quite comparable to observed orbits in Figure~\ref{YZs}. Obviously some unavoidable noise, present in the experimental version, lacks in the computations. This helps to realise that a fairly irregular  orbit, as in Figure~\ref{YZs}c, is in fact likely demonstrating the underlying presence of a periodic orbit, as in Figure~\ref{YZth}c.  

In fact, apart from an obvious difference in scaling, a reasonably good qualitative comparison  exists even with the simplified DLE
(\ref{eq:DLEQ}), see Figure~\ref{UDDth}. Of course,  it should be noted that in the DLE both the axes as well as the drive $\mathsf{w}$  are  dimensionless.  Both  can however be converted into dimensional terms. Based on the disk radius, acceleration of gravity and calibrated friction coefficient, the length  and time scales found in Section \ref{sec4} C are  $L=4.6$ mm  and $\tau=0.27$ s, respectively. This translates a dimensionless  distance  $z=10$ to  $Z=4.6$ cm, showing that the plots in Figure~\ref{UDDth} are quite comparable to those in both the lab as well as the simulations, shown in Figures~\ref{YZs} and \ref{YZth}, respectively. However, the values of the drive in the DLE were chosen for visual similarity  in Figure~\ref{UDDth}. When converted to dimensional values, $W =\mathsf{w} L/\tau$, these lead to  6.7, 7.8 and 12.3 cm/s, respectively. These are clearly higher than the drives used in the experiments in Figure~\ref{YZs}, no doubt related to the bold assumptions that went into the reduction of the actual equations to the DLE.

\begin{figure}[h]
\includegraphics[angle=0,width=\textwidth]{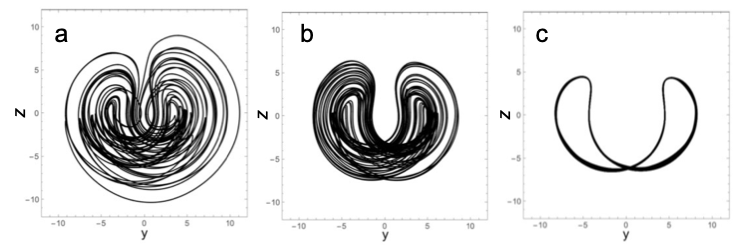}
  \caption{Numerically computed orbits of the DLE  (\ref{eq:DLEQ}) showing the $y,z$ coordinates in the vertical $\mathsf{y,z}$-plane for nondimensional upward-drive  (a) $\mathsf{w}=$ 3.9, (b) 4.5, (c) 7.1. }\lb{UDDth} 
\end{figure}

\subsection{Comparison in Three-dimensional Phase Space}

The nice comparison between observed and modelled center-of-mass motions in the vertical plane invites a comparison of three-dimensional orbits in full $\mathcal{J} YZ$ phase space, that embeds real $YZ$-space. Figure~\ref{YZphsp} shows the orbits observed in the upward-driven disk, whereas Figure~\ref{YZthphsp} displays the full orbits obtained from the tuned equations (\ref{eq:YZeqs}), combined with (\ref{angmom4}) and (\ref{kt}). The correspondence is remarkable, showing detailed similarities, especially for large drive. In the latter case, where the assumptions for using the simplified DLE (\ref{eq:DLEQ}) are met best, the predicted behavior models the disk's motion well. Recall that numerical integration of the DLE reveals that (\ref{eq:DLEQ}) has a rich bifurcation structure\cite{VanderSchrier2000,maas2004theory}, of which these figures display snapshots of a chaotic, high-periodic and low-periodic state respectively. A more detailed quantitative comparison of this bifurcation structure awaits further experiments.
The agreement found sofar between experimentally observed and numerically simulated orbits lends credibility to the use of both models, although for the DLE this is restricted to large drives. However, the model's inability to capture the experimentally observed rotating state motivates questioning its completeness, a question addressed in the next subsection.
\begin{figure}[!t]
\includegraphics[angle=0,width=\textwidth]{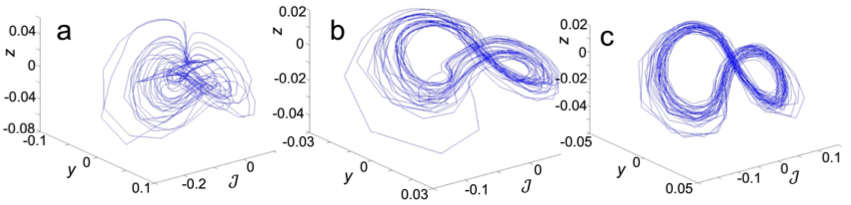}
 \caption{Phase space orbits $\mathcal{J} (m^2/s),Y(m),Z(m)$ for upward-driven disk at upward drive  $W$ equalling (a) 3.9, (b) 5.5 and (c) 7.1 cm/s.}\lb{YZphsp} 
\end{figure}

\begin{figure}[!ht]
\includegraphics[angle=0,width=\textwidth]{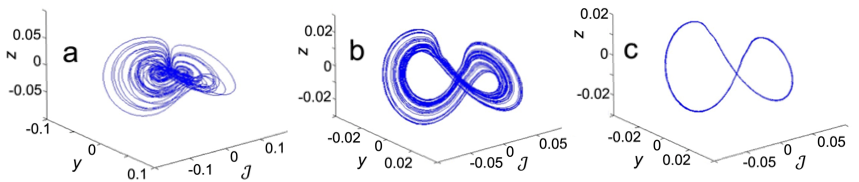}
\caption{Phase space orbits $\mathcal{J}(m^2s^{-1}),Y(m),Z(m)$ computed using the tuned disk equations at upward drive  $W$ equalling (a) 3.9, (b) 5.5 and (c) 7.1~ cm/s.}\lb{YZthphsp} 
\end{figure}
\subsection{Stability of Steadily Rotating Disk}  
We observed that for large enough drive  $W$, disk motions  follow either a periodic or chaotic orbit. These states apparently co-exist with states describing steady rotation of the disk, $\Omega=$ constant. These rotating states proved to be quite resilient. In a fluid such steadily rotating equilibria would be associated with convective motion. Their appearance in the upward-driven disk, a mechanical device, is  enigmatic. In a fluid, diffusion, produced by molecular exchange processes, is essential for the existence of these equilibria. And in the waterwheel these are present because the perforated cups leak water, enabling the center-of-mass to slowly creep inwards when no more water is added\citep{matson2007malkus,strogatz2018nonlinear}.  But  this possibility is absent in the theoretical model of the upward-driven disk, as will be shown here. In these rotating steady states, the assumptions underlying the simplification of the theoretical model to the dimensionless DLE are satisfied (that is, $R \ll \mathcal{R}$, while in  steady states,   friction factor $k$ is clearly constant). The theoretical absence of stable, steadily rotating states of the DLE, equations (\ref{eq:DLEQ}), can be inferred as follows. While equilibria of these equations (below, indicated by a bar), obtained by setting the time-derivative terms in (\ref{eq:DLEQ}) to zero, \textit{do} exist as steadily rotating states, 
\beq \bar{\omega}=\bar{y}=\pm \sqrt{\mathsf{w}}, ~\bar{z}=0, \lb{equi} \eeq
they are unstable to infinitesimal perturbations. 
To see this, perturb this equilibrium state,  assuming ${\bf x}\equiv( \omega,y,z)= \bar{\bf x} +{\bf x}'(t)$, the primed term being a perturbation. Inserting these in (\ref{eq:DLEQ}), linearising around the equilibrium state (\ref{equi}), and assuming the perturbations 
${\bf x}'={\bf x}_0 e^{\lambda t}$, where ${\bf x}_0$ represents an initial state close to equilibrium, we obtain a cubic eigenvalue equation for growth rate $\lambda$:
\beq \lambda^3 + \lambda^2 + \mathsf{w} \lambda + 2\mathsf{w}=0.\lb{ev1}\eeq
The steadily-rotating equilibria are stable when all eigenvalues have negative real part\cite{strogatz2018nonlinear}. But this is not the case as the numerically computed eigenvalues in Figure~\ref{rootsDLE} show. There is one negative real root and two complex conjugate roots having positive real part. In fact, even taking into account that the moment of inertia contains a time-varying component $\mathcal{I}_O$ and using the more realistic, state-dependent friction coefficient (\ref{kt}),  does not stabilize these steadily-rotating
 equilibria (not shown).  
\begin{figure}
\includegraphics[width=0.5\textwidth]{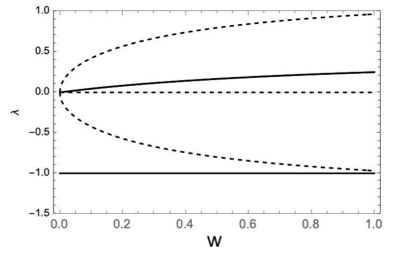}\lb{rootsDLE1}
\caption{Real (solid) and imaginary (dashed) parts of eigenvalues of matrix linearized around the steadily rotating equilibria (\ref{equi}).}
\lb{rootsDLE} 
\end{figure}

Stability is however achieved  by adding a diffusion-like damping term to the disk's vertical constraint equation (\ref{eq:DLEQ3}): \beq \dot{z} =-\omega y+\mathsf{w}- B  \mathsf{w} z,\lb{Zconstr} \eeq
where $B$ is taken constant. To keep the equations simple, we  retain their dimensionless form. 
\begin{figure}[t]
\includegraphics[width=.9\textwidth]{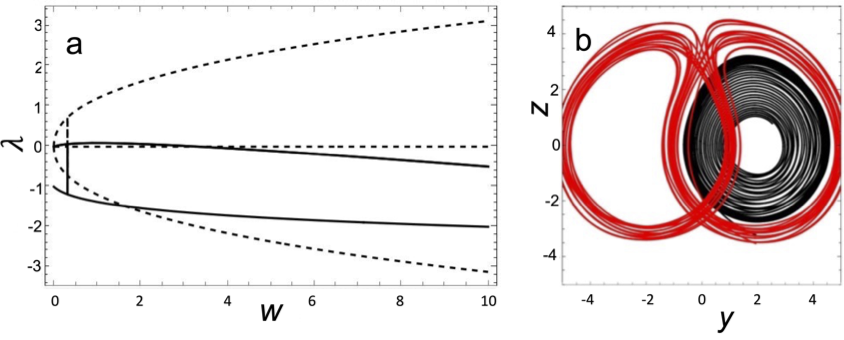}
\caption{(Color online) (a) Real (solid) and imaginary (dashed) parts of eigenvalues of matrix linearized around the steadily-rotating equilibria when a damping term is assumed in the vertical rolling constraint having damping coefficient  $B=0.2$; (b) Stable  orbit (black) winding inwards towards the steadily-rotating point at $\bar{\omega}=\bar{y}=\sqrt{\mathsf{w}}, \bar{z}=0$ for $\mathsf{w}=3.7$ together with a chaotic orbit (red). These orbits start in $\bar{\omega}=\bar{y}=\sqrt{\mathsf{w}}, \bar{z}=-3.2$ and -3.5, respectively.}
\lb{mult2} 
\end{figure}
Figure~\ref{mult2}a shows that in this case the steadily-rotating  equilibria turn stable to infinitesimal displacements above a certain minimal drive $\mathsf{w}_{min}$, which qualitatively corresponds with observations on the stability of the upward-driven disk mentioned in Section \ref{subsec3C}. Still, numerical integrations,  as in figure~\ref{mult2}b, show that  stability of these steadily-rotating  equilibria may actually extend to finite amplitude perturbations, occurring even below  $\mathsf{w}_{min}$. The steadily-rotating equilibria may be stable over the same parameter range as where chaotic or periodic states may occur. Their basins of attraction determine which state is reached given a certain initial state.

To motivate the addition of  damping  term$- B  \mathsf{w} z$ to (\ref{Zconstr})  note  that: (a) the full physical model, equations (\ref{kt},\ref{eq:YZeqs},\ref{angmom4}), captures  the observed disk behavior well (Subsections  A-C); (b) as the DLE, for large drives this model possesses two unstable rotating equilibria,  simultaneous with chaotic or periodic states; (c) experimentally these states are stable; (d)  adding this damping term to the  model renders these rotating states stable.  Thus, the credibility of the full physical, and for large drives, approximating DLE models  justifies looking for missing  physics that this term may represent. 

The added  term's  proportionality to $\mathsf{w}$ guarantees that at rest the disk will not move when the drive is off. Such  proportionality of damping to velocity is indeed known for some  materials (in particular rubber)\citep{cross2005increase,besson2007teach}.  
As this dynamic friction term also depends linearly on O's vertical position $z(t)$, it  implies that the net driving force, $\mathsf{w}(1-Bz)$, differs depending on whether the disk is pushing  or pulling  the wheels ---as when its center-of-mass lies above ($z>0$) or below ($z<0$) the point of contact O. It suggests the drive is strongest (for a  given motor setting) when the disk hangs at its most extreme location below O, while it is at its weakest in the opposite extreme. In the end, this leads to a hypothesis on the nature of friction acting on an object that is constrained to move  against the direction of gravity. Since much regarding friction is ill-understood at present\citep{hahner1998rubbing,besson2007teach,cross2016coulomb,pennestri2016review} (there is a whole field of physics, tribology, devoted to the study of friction), the physical nature of these frictional processes lying in the interlocking and deformation of roughness elements on the wheels and disk, this hypothesis  is now open for future experimental investigation. 

\section{Discussion}\lb{sec6}

The behaviour of the pendulum has been studied since  Mesopotamian and Egyptian times in the Third Millenium BC\cite{Boucher}. For small angle oscillations, it served as a clock\cite{baker2008pendulum}. Classically, the pendulum is  driven via the angular momentum equation (\ref{angmom4}) in one of three ways: (i) by adding a forcing term to its right-hand side,  in the form of  a periodic torque\cite{baker2008pendulum}; (ii) by oscillating its point of suspension, which, when done vertically, is expressed as a periodic modulation of the acceleration of gravity\cite{baker2008pendulum}; or (iii) by periodically adding energy while simultaneously reducing the moment of inertia through the pumping action of a swinger\cite{tea1968pumping}. 

While a truly frictionless pendulum would make an ideal clock, a periodically forced pendulum does not make a good substitution. Nonlinear oscillators that are periodically-forced near resonance, as pendulums\cite{baker2008pendulum} or  tidally-driven coastal basins\cite{maas1997nonlinear},  suffer  from natural detuning and display long-term modulation\cite{glendinning2020adaptive}. This is because the period of its response depends on its amplitude. Thus, the periodically-forced classical pendulum is unable to display a fixed period\cite{glendinning2020adaptive}. At some point in time, the regular forcing acts as a brake as it gets out-of-phase with the response. Historically, pendulum-based clocks were therefore  cleverly modified by adding semi-cycloid `cheeks'\cite{huygens1986pendulum}. During a pendulum's swing this guarantees that its length changes periodically, such that it stays isochronous, regardless of its amplitude\cite{huygens1986pendulum,baker2008pendulum}. 

In contrast to the classically forced pendulums, a steady upward drive has no frequency, hence cannot resonate and should thus not suffer from natural detuning nor exhibit long-term modulation.  Since  for large drives the upward-driven disk indeed obtains a well-defined period \citep{VanderSchrier2000}, it appears to be  a naturally-tuned isochronous pendulum. It is  robust to perturbations and is characterized by  its center-of-mass tracing a bent figure-eight type of motion, see figures~\ref{YZs}c
 and \ref{UDDth}c. Its orbit resembles  that of the centre-of-mass of a person pumping a swing\cite{tea1968pumping}. In this regime, the steadily-forced circular pendulum acts as a DC-AC converter, which \emph{does} meet the ancient goal of constructing a clock with a well-defined unique period.

This upward-driven disk is easy to manufacture and can thus serve to illustrate sensitive dependence on initial conditions, chaos and multiple equilibria in the class-room. As the disk's behaviour is mimicked to a large extent by solutions of the one-parameter forced-and-damped physical pendulum ---a simplified version of the  Lorenz equations--- this allows connecting the mechanical model to a simplified mathematical representation suitable for further experimental investigation of symmetry-breaking and period-doubling bifurcations\citep{VanderSchrier2000}. 

This correspondence, in turn, suggests the upward-driven disk also offers a purely mechanical analog for convection in a fluid, interpreting convection simply as driving the center-of-mass of a density-stratified fluid upwards towards instability. When the fluid's center-of-mass rises above the geometric center of the fluid-basin, the fluid will overturn. The rate at which overturning sets in, and the maximum height to which the center-of-mass is able to grow, depends on the off-centre distance $y(t)$ at which it passes the geometric center during its ascent.  

The disk also displays two stable, steadily-rotating states, not present in (\ref{eq:DLEQ}). To obtain such a state, this mathematical model requires  modification.  This is unexpectedly achieved by extending  the rolling constraint in vertical direction. Physically, it would mean  that the force with which the wheels squeez the disk while pushing it upwards seems to depend on whether the disk pushes or pulls on the rotating wheels, as when its center-of-mass is above or below the  point where it is sandwiched, possibly a consequence of the time-variable contribution of gravity to the force of constraint.
 
\begin{acknowledgments}
The author is much indebted to Gerard van der Schrier for inspiring discussions leading to the idea behind the device, to Dirk-Jurjen Buijsman for suggesting the use of a disk and for the  construction of the LEGO version,  to Frans Eijgenraam and Jasper Witte for assistance with the video data acquisition and processing, and to Huib de Swart for discussions and comments on an earlier draft. Perceptive and constructive comments by two anonymous reviewers and reviewer Kirk McDonald are gratefully acknowledged.\\

\end{acknowledgments}

\section{Author Declarations}
The author has no conflicts to disclose.
\bibliographystyle{natbib}
\bibliography{bibliography}
\end{document}